\documentclass[a4paper]{article}
\usepackage{lineno}
\usepackage[utf8]{inputenc}
\usepackage[a4paper,top=3cm,bottom=2cm,left=3cm,right=3cm,marginparwidth=1.75cm]{geometry}

\usepackage{amsmath,amssymb} 
\usepackage{color}
\usepackage{amsmath}
\usepackage{amssymb}
\usepackage{algorithm}
\usepackage{algorithmicx}
\usepackage{algpseudocode}
\usepackage{algpascal}

\usepackage{mathrsfs}
\usepackage{soul}
\usepackage{multirow}
\usepackage{comment}
\usepackage{xfrac}
\usepackage{rotating}
\usepackage{hhline}
\usepackage{mathtools}
\usepackage{wrapfig}
\usepackage{upgreek}
\usepackage[colorlinks=true]{hyperref}
\usepackage{tikz}
\usepackage{color, colortbl}
\usepackage[nomarkers]{endfloat}
\definecolor{Gray}{gray}{0.9}
\usepackage{booktabs}
\usepackage[utf8]{inputenc}
\usepackage[misc]{ifsym}
\usepackage{lipsum}
\usepackage{sidecap}
\usepackage{subcaption}
\usepackage{authblk}
\usepackage{pgfplots}
\usepackage{pgfplotstable}
\usepackage{filecontents,pgfplots}
\usetikzlibrary{arrows,calc,shapes,snakes,positioning,matrix,arrows,decorations.pathmorphing,decorations.text}
\usepackage{sci}

\newcommand{\ourMethodNoSpc}{BrainSynth}
\newcommand{\ourMethod}{\ourMethodNoSpc~}

\newcommand\blfootnote[1]{%
  \begingroup
  \renewcommand\thefootnote{}\footnote{#1}%
  \addtocounter{footnote}{-1}%
  \endgroup
}

\title{Metadata-Conditioned Generative Models to Synthesize Anatomically-Plausible 3D Brain MRIs}
\author[1]{Wei Peng}
\author[2]{Tomas Bosschieter}
\author[3]{Jiahong Ouyang}
\author[4]{Robert Paul}
\author[1,5]{Ehsan Adeli}
\author[6, \Letter]{Qingyu Zhao}
\author[1,7, \Letter]{Kilian M. Pohl}
\affil[1]{{\small Department of Psychiatry \& Behavioral Sciences, Stanford University, Stanford, CA 94305}}
\affil[2]{{\small Institute for Computational and Mathematical Engineering, Stanford University, Stanford, CA 94305}}
\affil[3]{{\small Department of Electrical Engineering, Stanford University, Stanford, CA 94305}}
\affil[4]{{\small Missouri Institute of Mental Health,  University of Missouri, St. Louis, MO 63121}}
\affil[5]{{\small Department of Computer Science, Stanford University, Stanford, CA 94305}}
\affil[6]{{\small Department of Radiology, Weill Cornell Medicine, New York, NY 10065}}
\affil[7]{{\small Center for Health Sciences, SRI International, Menlo Park, CA 94025}}

\begin{document}
\maketitle

\begin{abstract}
Generative AI models hold great potential in creating synthetic brain MRIs that advance neuroimaging studies by, for example, enriching data diversity.  However, the mainstay of AI research only focuses on optimizing the visual quality (such as signal-to-noise ratio) of the synthetic MRIs while lacking insights into their relevance to neuroscience.  To gain these insights with respect to T1-weighted MRIs, we first propose a new generative model, BrainSynth, to synthesize metadata-conditioned (e.g., age- and sex-specific) MRIs that achieve state-of-the-art visual quality. We then extend our evaluation with a novel procedure to quantify anatomical plausibility, i.e., how well the synthetic MRIs capture macrostructural properties of brain regions, and how accurately they encode the effects of age and sex. Results indicate that more than half of the brain regions in our synthetic MRIs are anatomically accurate, i.e., with a small effect size between real and synthetic MRIs. Moreover, the anatomical plausibility varies across cortical regions according to their geometric complexity. As is, our synthetic MRIs can significantly improve the training of a Convolutional Neural Network to identify accelerated aging effects in an independent study. These results highlight the opportunities of using generative AI to aid neuroimaging research and point to areas for further improvement. \blfootnote{\Letter \space Correspondence to \href{qiz4006@med.cornell.edu}{Qingyu Zhao} and \href{kilian.pohl@stanford.edu}{Kilian M. Pohl}}

\end{abstract}

\section{Introduction}
Generative models are designed to learn the underlying patterns and structure of the training data and use that knowledge to generate new samples that resemble the original data~\cite{goodfellow2020generative,karras2019style}. Recently, these models have significantly impacted a wide range of applications, such as image synthesis (e.g., artwork creation~\cite{karras2019style}), natural language processing (e.g., chatbots~\cite{bhirud2019literature}), and healthcare (e.g., medical report generation~\cite{kisilev2015medical} and drug discovery~\cite{zeng2022deep}). Furthermore, their power to create a large amount of ``novel content" has particular relevance to clinical neuroimaging investigations, which have historically relied on relatively small datasets due to challenges associated with acquisition of high quality scans (e.g.,  due to costs, quality control concerns, access to particular clinical populations). \cite{szucs2020sample}. 

Despite this compelling vision, existing attempts in synthesizing brain MRIs are mainly proof-of-concepts \cite{sun2022hierarchical,hong20213d} focusing on translating successful models in computer vision~\cite{ho2020denoising} to neuroimaging. This translation is challenging as early generative models ~\cite{kwon2019generation,sun2022hierarchical} (mostly based on Generative Adversarial Networks (GANs)~\cite{goodfellow2020generative}) can only synthesize relatively low-resolution 2D natural images. Given that the number of voxels in a typical T1-weighted brain MRI (with $1mm$ resolution) is similar to the number of pixels in a 4K image (i.e., ultra-high definition), those models generate low-quality MRIs that are very noisy or blurry visualizations of the brain. Synthesizing high-quality MRI is difficult for GANs as their challenging min-max optimization cannot stably capture the distribution of high-resolution data due to model collapse \cite{saxena2021generative}. A more stable alternative to GANs is diffusion models~\cite{ho2020denoising}, which directly estimate data distributions and provide fine-grained control in the synthesis process 
~\cite{ho2020denoising,rombach2022high}. Hence, the diffusion model by Pinaya et al.~\cite{pinaya2022brain} synthesized T1-weighted brain MRIs of superior image quality to those produced by GANs according to subjective visual assessments (100,000 synthetic MRIs are publicly available) and metrics popular in computer vision (e.g., signal-to-noise ratio).
To efficiently synthesize MRIs of even higher image quality, we propose a diffusion model, called \ourMethodNoSpc, that couples feature quantization~\cite{esser2021taming} with a fine-grained coding technique explicitly accounting for metadata information (i.e., age and sex).


While high image quality is a prerequisite for a meaningful followup investigation, the core of most clinical neuroscience studies is to study the anatomical properties of individual brain regions, such as identifying regional atrophy caused by aging and cognitive impairment \cite{raz2005regional,morrison2021regional}. To align with neuroscience, we thus argue that a key analysis missing from existing generative models is the assessment of anatomical plausibility, i.e., how well they capture those anatomical properties. Hence, we propose a framework for doing so, which reveals that the anatomical plausibility of our synthesis is linked to the geometric complexity of cortical regions. We also use this evaluation framework to quantify the anatomical plausibility in synthesizing  age- and sex-specific brain MRIs. Finally,  we discuss the strengths and limitations of using these synthetic MRIs to improve the sensitivity in identifying accelerated aging effects linked to cognitive impairment. 


%
%
\section{Results}
\subsection{Synthesizing Brain MRIs via Deep Generative Models}
\begin{figure}[!t]
    \centering
    \includegraphics[width=1\textwidth]{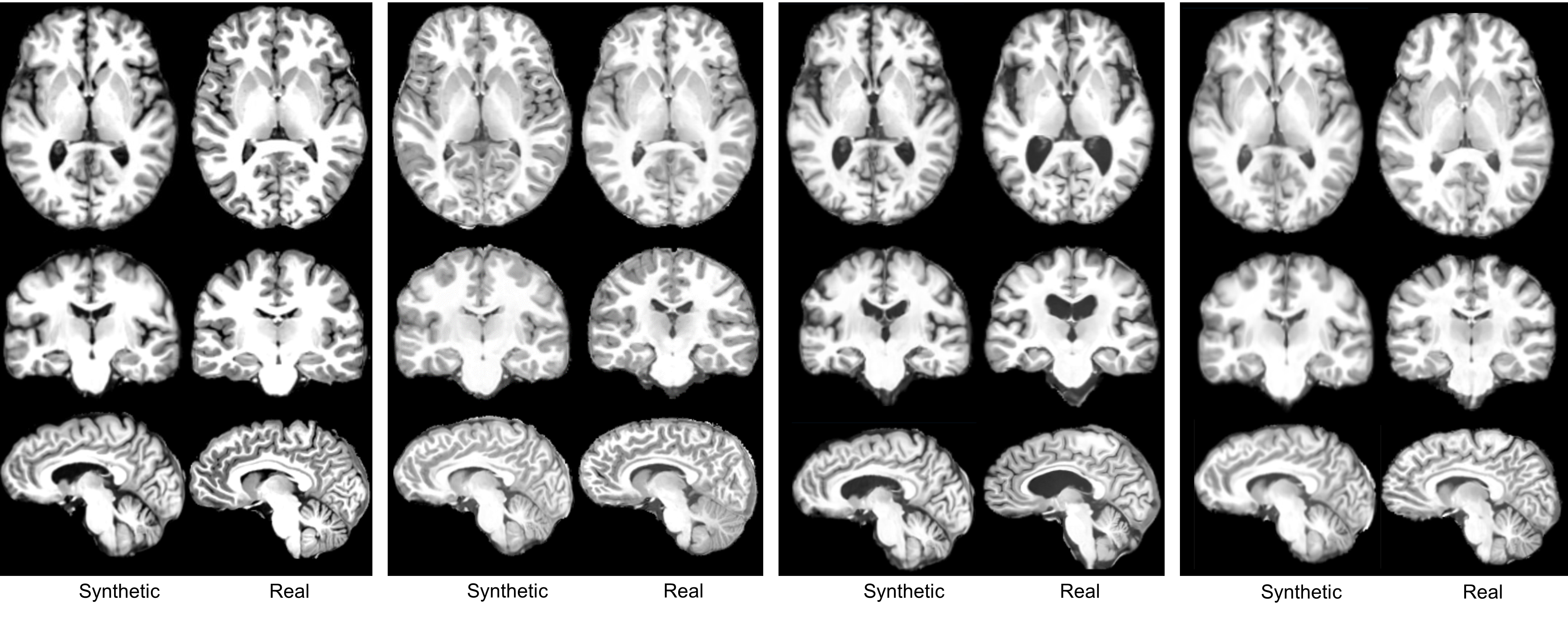}
    \caption{The axial, coronal, and saggital view of 4 synthetic T1-weighted MRIs generated by \ourMethod and their closest real MRIs among the training data according to the L2 distance between their latent representations.}
    \vspace*{6.5in}
    \label{fig:realvsSyn}
\end{figure}

\ourMethod was trained on 3996 of 4296 preprocessed T1-weighted MRIs of normal controls pooled from 3 neuroimaging studies (see Table \ref{tab:demographics}): the Alzheimer's Disease Neuroimaging Initiative~\cite{petersen2010alzheimer} (ADNI), the National Consortium on Alcohol and Neurodevelopment in Adolescence~\cite{brown2015national} (NCANDA), and an in-house dataset from SRI International~\cite{zhao2021longitudinal} (SRI). Each scan was denoised, homogeineity-corrected, skull-stripped, and affinely aligned to a template (see Section \ref{sec:materials}). After training, our model generated 10,000 new independent synthetic brain MRIs. Each MRI was constructed by first randomly selecting a sex and an age from 13 to 91 years. Conditioned on sex and age, our method then generated a lower-dimensional latent representation and finally decoded the latent representation to create the MRI (see Section \ref{method}). Figure~\ref{fig:realvsSyn} shows four examples of the synthetic MRIs and their ``closest looking" real MRIs in the training dataset measured by the Euclidean distance (L2 norm) in the latent space. All 10,000 synthetic MRIs were `new' samples, i.e., they differed from the training data as 1) the L2 norm between each synthetic MRI and its closest MRI was significantly greater than 0 ($p<0.001$, 6890.53 $\pm$ 4536.80) and 2) none of the real and synthetic MRIs had identical latent representations.

\begin{table}[t!]
    \caption{Demographic of four datasets. The first three datasets consisted of 4296 T1-weighted MRIs (from 1,236 normal controls), which were used for training \ourMethod and assess the quality of synthesis. A CNN predicting age from T1-weighted MRIs was trained on the UCSF dataset, which consisted of normal controls, People Living with HIV (PLWH), Mild Cognitive Impairment (MCI) and HIV-associated Neurocognitive Disorder (HAND). Sex is specified by the ratio between males and females. Age is itemized by the mean $\pm$ `standard deviation' [ age range ] of the cohort.    }
    \label{tab:demographics}
    \centering
    \begin{tabular}{l >{\centering\arraybackslash}p{0.19\textwidth} >{\centering\arraybackslash}p{0.13\textwidth} >{\centering\arraybackslash}p{0.12\textwidth} >{\centering\arraybackslash}p{0.09\textwidth}>{\centering\arraybackslash}p{0.19\textwidth}}
    
        \hline
        \multirow{2}{*}{\textbf{Dataset}}
         & \multirow{2}{*}{\textbf{Cohorts}}  &  \textbf{Number of}  & \textbf{Number of}  & \textbf{Sex} & \textbf{Age} \\
        &  & \textbf{Scans} &  \textbf{Subjects} & (M/F) & (Years)   \\
        \hhline{======}
        ADNI & Normal Control  & 1215 & 342 & 160/182 & $76.30\pm 6.04 $ ~[59, 91] \\
        \hline
        NCANDA & Normal Control & 2285 & 621 & 300/321 & $18.34\pm 2.73$ ~[13, 27] \\
        \hline
        SRI & Normal Control & 796 & 273 & 129/144 & $48.52\pm 19.9$ ~[19, 86] \\
        \hhline{======}
        \multirow{4}{*}{UCSF} & Normal Control & 156 & 156 & 146/10 &$ 69.06\pm 5.43$~[55, 78] \\
                             & PLWH   & 37   & 37 & 36/1 & $64.18\pm 3.50$  ~[59, 75]\\
                             & MCI & 148 & 148    & 71/77 & $66.09\pm 6.74$  ~[55, 80]\\
                             & HAND & 145 & 145   & 136/9 & $63.45\pm 4.63$  ~[55, 78]\\
        \hhline{======}
    \end{tabular}
    \vspace*{6.5in}
\end{table}

    

Compared to MRIs synthesized by traditional generative models~\cite{MultiContrastGAN2019,yu20183d,shin2018medical} (e.g., those based on generative adversarial networks), the MRIs generated by \ourMethod were visually more similar to real MRIs (Supplement Fig. S1), had significantly smaller discrepancy (Maximum-Mean Discrepancy Score) from real MRIs ($p<0.001$, two-sample $t$-test), and had significantly higher Signal-to-Noise Ratio ($p<0.001$, two-sample $t$-test) (Table \ref{tab:comparison}). Furthermore, only the synthesized MRIs of our model resulted in a 2D distribution (by applying t-SNE~\cite{van2008visualizing} to their latent representations) that matched those of the real MRI (Supplement Fig. S1). Based on these findings, we focus the remainder of this section on the synthetic MRIs of \ourMethodNoSpc.

\begin{table}[!t]
    \centering
    \caption{Image quality of 300 synthetic MRIs generated by \ourMethod and comparison methods. \ourMethod records the best scores (typeset in bold) among all methods with the MS-SSIM score being closest to the real MRIs (MS-SSIM: 0.931) of the replication set, the lowest MMD, and the highest SNR.}
    \label{tab:comparison}
    \begin{tabular}{r|c|c|c}
    &  MS-SSIM & MMD $\downarrow$ & SNR $\uparrow$\\
        \hline \hline
    VAE-GAN~\cite{larsen2016autoencoding} &0.803  &    84561.61   &    5.101     \\
    CCE-GAN~\cite{xing2021cycle} &   0.816 &   38712.64   &   7.217    \\
    $\alpha$-WGAN~\cite{kwon2019generation} & 0.832 & 11459.86    &   6.821  \\
    3D-DPM~\cite{dorjsembe2022three} &     0.876      &  11465.11   &   13.842   \\
    Latent~\cite{pinaya2022brain} &     0.919      &   9614.704   &   15.509   \\
    \textbf{\ourMethod (proposed)} & \textbf{0.933}&  \textbf{5114.63} & \textbf{18.223}  \\
    \end{tabular}
    \vspace*{7.5in}
\end{table}

\subsection{Accuracy of Anatomical Structures in Synthesized MRIs}
\begin{figure}
    \centering
    \includegraphics[width=1\textwidth]{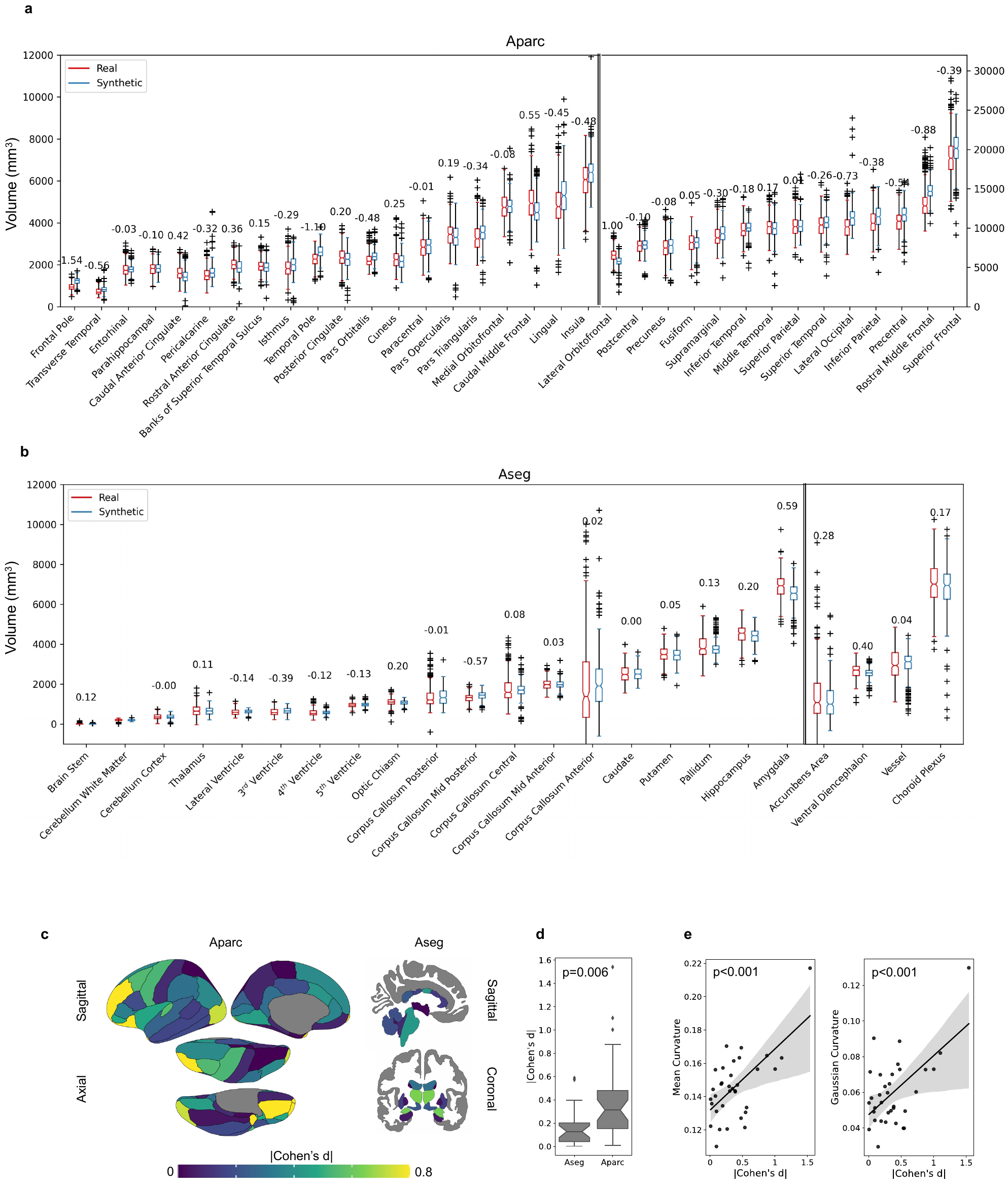}
    \caption{Distribution of the volume for each of the (a) 34 cortical regions (i.e., Aparc~\cite{desikan2006automated}) and (b) 23 subcortical regions (i.e., Aseg~\cite{fischl2002whole}) in 300 real and 300 age- and sex-matched synthetic MRIs. For each region, the number above the boxplot is the `Cohen's d' between the distributions of real and synthetic MRIs, with values closer to 0 indicating more overlap between them. With respect to the Aparc regions, the magnitude of Cohen's d (color-coded in (c)) was significantly different from that of Aseg regions (d) and correlated with mean and Gaussian curvature (e). }
    \label{fig:Acc4Anatomical}
     \vspace*{6.5in}
\end{figure}

To quantify the anatomical plausability of the synthesis, we created a replication set consisting of 300 real MRIs that were randomly selected from the three data sets (92 from ADNI, 152 from NCANDA, 56 from SRI), evenly distributed with respect to sex and age between 13 to 91 years, and not used for training (see Section \ref{sec:materials}). 300 synthetic MRIs with the same sex and age distribution were randomly selected from the 10,000 data set. For each MRI (synthetic or real), the Freesurfer pipeline~\cite{fischl2012freesurfer} estimated the volume of 34 cortical regions defined by the Desikan-Killiany atlas~\cite{desikan2006automated} (a.k.a., Aparc) and 23 subcortical regions~\cite{fischl2002whole} (a.k.a., Aseg). For each region, we quantified the difference in distribution between the real and synthetic data set (see Fig \ref{fig:Acc4Anatomical}a$\&$b) by computing the effect size (Cohen's $d$ \cite{cohen2013statistical}). Note, we did not test for statistical significance for those group differences as $p$-values can be made arbitrarily small by sampling an infinite number of synthetic MRIs. 29 out of the 57 regions (51\%) were associated with small effect sizes ($|d|<0.2$), 19 regions (33\%) with small-to-medium effect sizes ($|d|<0.5$), and 9 regions (16\%) with medium-to-large effect sizes ($|d|\geq0.5$) (Fig. \ref{fig:Acc4Anatomical}c). The magnitude of effect size $|d|$ associated with Aseg regions was significantly lower than that of Aparc regions (Fig. \ref{fig:Acc4Anatomical}d, two-tailed $p=0.006$, $t$-test). 

With respect to the 34 Aparc regions,  $|d|$ significantly correlated with the (population-averaged) mean curvature and Gaussian curvature of those regions (two-tailed $p<0.001$, Fig. \ref{fig:Acc4Anatomical}e) but not with thickness and surface area. Lastly, $|d|$ did not significantly differ across 4 major lobes defined by the Desikan-Killiany atlas ($p=0.38$, ANOVA).

\subsection{Encoding Sex Differences and Aging Effects}
\begin{figure}
    \centering
    \includegraphics[width=1\textwidth]{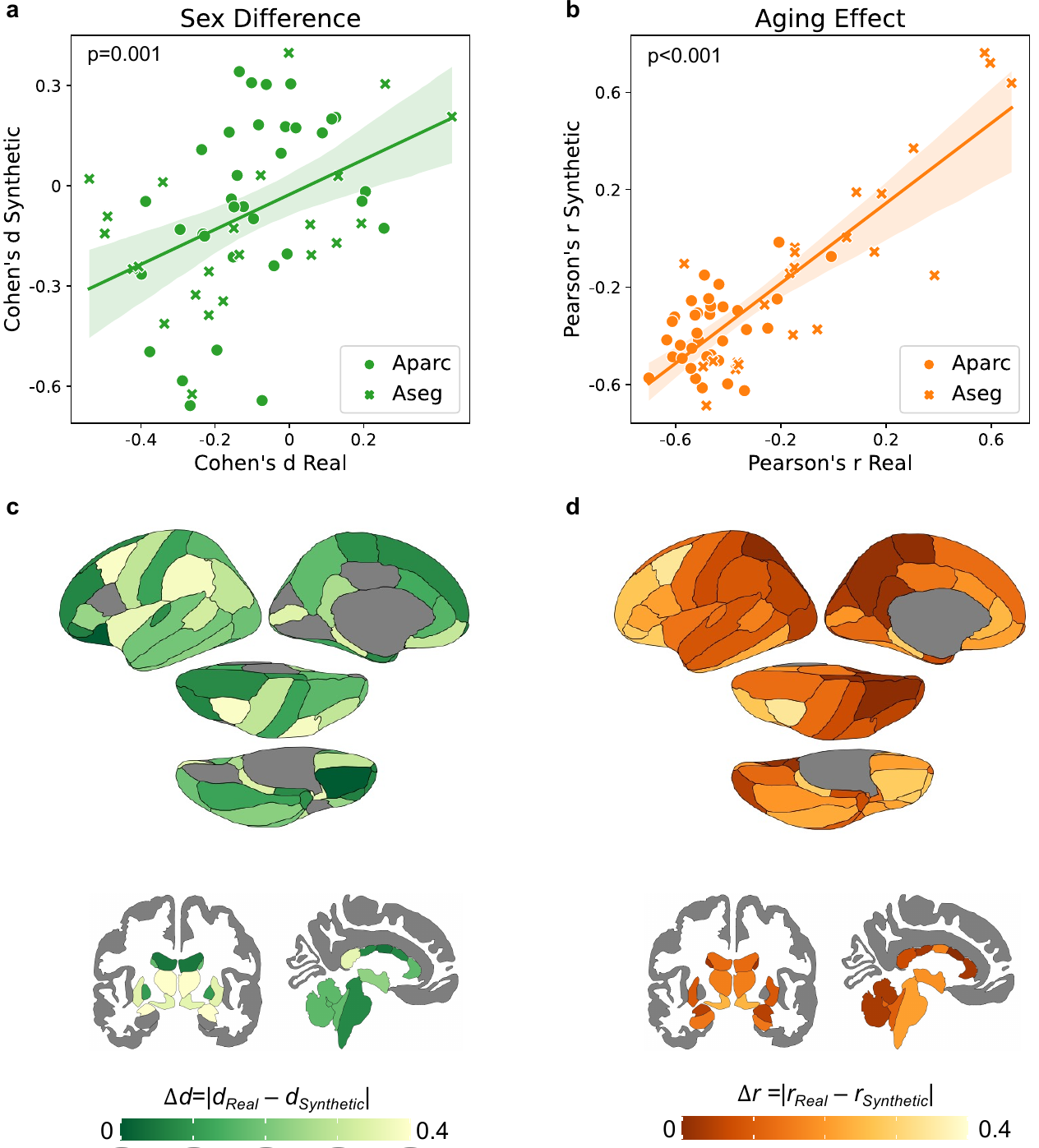}
    \caption{(a) Sex differences (Cohen's d between regional volume measurements of males and females) and (b) aging effects (Pearson's r between volume and age) observed in synthetic MRIs significantly correlated with those measured in real MRIs. Brain regions are color-coded according to the magnitude of sex difference (c) and aging effects (d).}
    \label{fig:SexAgingEffects}
     \vspace*{6.5in}
\end{figure}

Using the regional volume scores from the prior experiment, we examined how similar the aging effects and sex differences captured by the synthetic MRIs were to those observed in real MRI. With respect to quantifying sex differences, we computed the effect size (i.e.,  Cohen's $d$) between the 150 males and 150 females for each regional volume measurement and data set (Supplement Fig. S2). The resulting effect sizes across all regions significantly correlated between the two data sets ($r=0.421$, $p=0.001$, Fig. \ref{fig:SexAgingEffects}a). To further quantify the accuracy of the `synthesized' sex difference, we computed for each region the absolute difference $\Delta d$ between the effect sizes of the two data sets (Fig. \ref{fig:SexAgingEffects}c). $\Delta d$ did not significantly differ between the Aparc and Aseg regions ($p=0.95$, two-sample $t$-test). $\Delta d$ also was insignificantly different across lobes ($p=0.203$, ANOVA) and did not correlate with morphological properties (curvature, thickness, area) across different Aparc regions ($p>0.2$). 
To measure the accuracy of the `synthesized' aging effect, we separately computed Pearson's $r$ between the volume measurement and age for each of the 57 regions and data sets (Supplement Figs. S3$\&$S4). Aging effects estimated on synthetic data significantly correlated with real aging effects across the 57 brain regions ($r=0.843$, $p<0.001$,  Fig. \ref{fig:SexAgingEffects}b). This correlation (Fig. \ref{fig:SexAgingEffects}b) was significantly greater ($p<0.001$, $z$-test) than the correlation between real and synthetic sex differences (Fig. \ref{fig:SexAgingEffects}a). Although the absolute difference $\Delta r$ between real and synthetic aging effects did not differ between Aparc and Aseg regions, $\Delta r$ was significantly higher in the frontal lobe than the temporal, parietal, and occipital lobes ($p=0.004$, ANOVA, Fig. \ref{fig:SexAgingEffects}d).


\subsection{Identifying Accelerated Aging Effects in an Independent Study}
\begin{figure}
    \centering
    \includegraphics[width=1\textwidth]{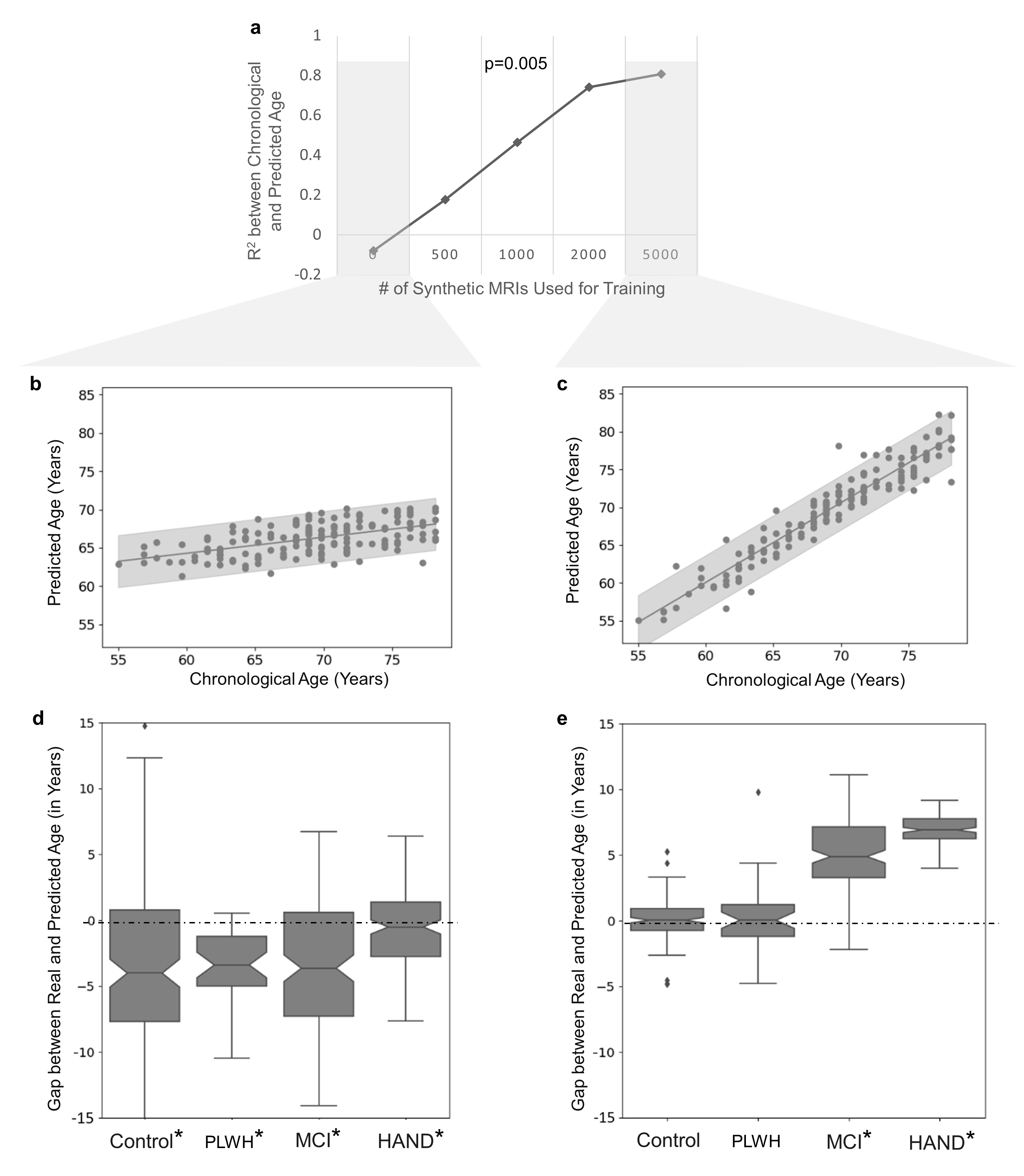}
    \caption{(a) When training a Convolutional Neural Network (CNN) to predict age from MRIs of normal controls, the prediction accuracy significantly increased with the number of synthetic MRIs being added to the training data set; (b) True chronological age vs predicted age by a CNN trained without using synthetic MRIs or (c) with 5000 synthetic MRIs; (d) When applying the trained CNN to predict age of controls and different patient groups, the CNN trained without synthetic MRIs resulted in predicted age significantly lower than true chronological age, i.e., negative brain-age gaps. * corresponds to $p<0.05$ in one-sample $t$-test; (e)
The CNN trained with 5,000 synthetic MRIs revealed accelerated aging effects in the two cohorts with cognitive impairment, i.e., with significantly positive brain-age gaps. }
    \label{fig:IndependentStudy}
     \vspace*{6.5in}
\end{figure}

Lastly, we investigated whether the synthetic MRIs of \ourMethod could be used to help in the training of a predictor of `brain age' (i.e., assessing the health of the brain) on an independent sample of 486 MRIs (ages 55 - 80 years) acquired from the UCSF Weill Institute for Neurosciences (PI: Victor Valcour) (see Table \ref{tab:demographics}, \cite{zhang2022multi}). This dataset consisted of 156 normal controls, 37 people living with HIV (PLWH), 148 older adults diagnosed with Mild Cognitive Impairment (MCI), and 145 diagnosed with HIV-associated Neurocognitive Disorder (HAND). The baseline predictor consisted of a standard Convolutional Neural Network (CNN)~\cite{hara2018can} trained and tested on the MRIs of the normal control cohort. 2-fold cross-validation resulted in a Mean Absolute Error (MAE) of 14.99 years and an $R^2$ of -0.08 (Fig. \ref{fig:IndependentStudy}b), which was not significantly better than chance. We also applied this predictor (trained on controls) to estimate the age of the three remaining cohorts. Like for controls, the ``brain-age gap" (i.e., the difference between brain and chronological age \cite{cumplido2023biological,ballester2023gray}) was significantly less than 0 for each of these cohorts ($p<0.015$; Fig.  \ref{fig:IndependentStudy}d).



Next, we repeated the 2-fold cross-validation but now included up to 5000 synthetic MRIs from the same age range to each training fold. The accuracy significantly increased with the number of added synthetic MRIs (p$=0.005$, Fig.  \ref{fig:IndependentStudy}a). Adding 5,000 synthetic MRIs to the training fold resulted in MAE=4.49 years and $R^2$=0.80 (Fig.  \ref{fig:IndependentStudy}c). Estimating brain age with this predictor (i.e., including 5000 augmented MRIs in the training set) resulted in the brain age of both controls and PLWH not to be significantly different from chronological age ($p>0.35$, Fig.  \ref{fig:IndependentStudy}e). The brain-age gap for MCI and HAND participants was significantly greater than 0, i.e., the estimated age was significantly older than the chronological age ($p<0.001$). These findings were confirmed when repeating the evaluation just using males (only MAE changed to MAE=4.45 years).

\section{Discussion}

We proposed a novel diffusion model for synthesizing realistic-looking brain T1-weighted MRIs that were visually superior and received better image quality scores compared to those synthesized by other state-of-the-art generative models. Our study was unique in quantifying the anatomical plausibility of the synthesis and the capability of capturing aging effects and sex differences in synthetic MRIs. We further showed the potential of the synthetic MRIs generated by our model in improving the generalizability of an MRI-based age predictor.

Synthesizing high-quality brain MRIs is a long-standing problem that arose even before deep generative models received attention \cite{collins1998design}. Early solutions are largely based on distorting a template MRI by simulating (plausible) tissue deformation \cite{xue2006simulating}. Given their high dependency on the template, the resulting MRIs are not independent samples and thus cannot increase data diversity. One way of increasing data diversity is by creating new independent MRIs from scratch, which is the purview of generative models. However, research along this direction is in its early stages as studies mostly investigate the feasibility of various methodologies~\cite{goodfellow2020generative,ho2020denoising}. Feasibility is generally defined with respect to visual authenticity and popular metrics applied to the entire image, such as SSIM~\cite{hong20213d,pinaya2022brain,xing2021cycle} and signal-to-noise ratio~\cite{sui2021mri}. Based on these assessments, \ourMethod generates MRIs of higher quality compared to other state-of-the-art generative models \cite{gulrajani2017improved,larsen2016autoencoding,pinaya2022brain,xing2021cycle,dorjsembe2022three,kwon2019generation}, i.e., with greater anatomical detail, lower signal-to-noise ratio, and higher visual authenticity (see Fig.~\ref{fig:realvsSyn} $\&$ S1). 

Missing from those assessments is anatomical plausibility, which is essential for the synthesized MRIs to substantially impact neuroimaging research, such as Brain-Wide Association Studies (BWAS)~\cite{marek2022reproducible}. Accordingly, we evaluated our generative model with respect to regional measurements computed by Freesurfer, which many BWAS studies use for their analysis. The regional measurements extracted from our synthetic MRIs were moderately accurate with the distributions of more than half of the brain regions highly overlapping with those from real MRIs (Fig. \ref{fig:Acc4Anatomical}). However, 16\% of regions had volumes substantially deviating from those observed in real data. The difficulty of synthesis in these regions was most likely driven by the geometric complexity of the cortex as the synthesis in the cortex was generally less accurate (higher Cohen's d) than for sub-cortical regions and the lowest accuracy was recorded in the cortical regions that were most curved (such as frontal pole). The findings from our evaluation point to areas of improvement (e.g., prefrontal cortex) and suggest that explicitly modeling geometric complexity could further advance the synthesis of MRIs. 

In addition to optimizing the quality of the synthesis, how to utilize those synthetic MRIs in followup investigations is still unclear. Notably, neuroimaging studies still largely rely on traditional group-level statistical tests that do not allow for an arbitrary inflation of sample size by synthetic MRIs. However, we showed that the quality of our synthetically generated MRIs was high enough to aid machine learning analysis to identify abnormal aging trajectories caused by neurological diseases. When using a machine learning model to predict age from MRIs~\cite{cole2017predicting}, the accuracy and robustness of these machine learning models (especially deep learning models) generally increase with the size of the training set~\cite{shorten2019survey}. It thus was not surprising that adding our synthetic MRIs to a training set helped in creating a more accurate predictor of age from MRIs of normal controls. The more accurate estimation resulted in increased sensitivity in measuring the brain-age gap, i.e., the difference between the estimated age based on the MRI and the true chronological age. Specifically, the age predictor trained with synthetic MRIs detected a positive age gap in patients diagnosed with MCI or HAND. This gap is widely viewed in neuroimaging research as a quantitative marker for accelerated brain aging \cite{cumplido2023biological,ballester2023gray,zhao2021longitudinal}. Further confirming the sensitivity of the predictor was a non-significant brain-age gap for individuals without cognitive impairment. A limitation in this analysis was its training was confined on synthetic and real MRIs representing normal controls so that any effect of a disease was simply viewed as accelerated aging. The capability of directly generating MRIs reflecting disease effects is a much more challenging task due to the difficulty of gathering a large enough number of MRIs acquired of patients to train the model in the first place. While one can pool a relatively large number of normal controls from multiple studies, it is generally impossible to do so for disease populations especially those that are harder to recruit (such as HAND). 

Another advantage of our model is that it can generate age- and sex-dependent MRIs. Generating metadata-dependent MRIs can be crucial for augmenting training data (as it was the case in our brain age experiment) as the synthetic MRIs can be tailored to a specific demographic construct of an independent study of interest. As such, the model might also be used to enrich the data of underrepresented samples in a study. Furthermore, our model could potentially become a novel way of understanding the biological effects of different factors (e.g., age) on brain morphometry. Traditional machine learning approaches for identifying aging effects largely rely on first training supervised models to predict age and then interpreting the learned model in a post-hoc manner, e.g., via saliency visualization \cite{zhao2019confounder}. In contrast, our model can directly synthesize MRIs of different ages, making it possible to generate age-dependent brain templates~\cite{dalca2019learning} that provide an intuitive visualization of the group-level aging effects. Lastly, our model can be extended to create ``counterfactual" MRIs specific to an individual~\cite{pawlowski2020deep}, e.g., by synthesizing the MRI of an individual at a specific age in the future (see additional analysis in Supplement Fig. S5). Such individualized counterfactual synthesis could be used to impute missing MRIs in longitudinal studies~\cite{matta2018making} and to characterize heterogeneity in subject-specific aging trajectories. 

%
%
\section{Methods}\label{method}
\begin{figure}
    \centering
    \includegraphics[width=1.1\textwidth]{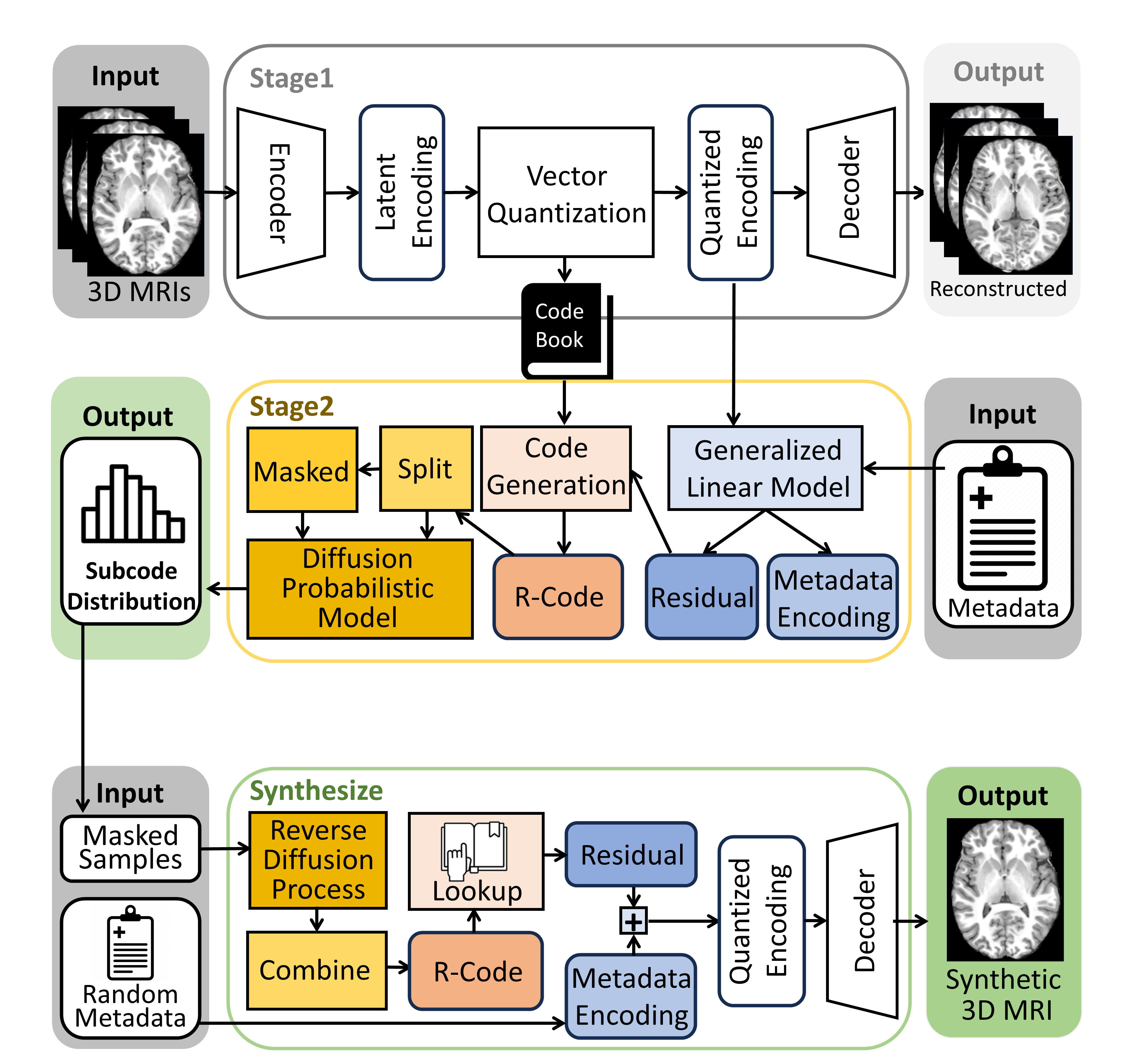}
    \caption{Architecture of \ourMethodNoSpc: Stage I consists of a VQ-VAE to reduce the real 3D MRI to a quantized vector representation derived from a code book. In stage II, the quantized code is disentangled into a metadata-related encoding (e.g., age and sex) and a residual (metadata-free). The distribution of the R-code (code of the residual) is then learned by our masked Diffusion Probabilistic Model. Once training is completed, MRIs are generated by combining (random) R-Code with metadata.}
    \label{fig:method}
     \vspace*{6.5in}
\end{figure}

\ourMethod is a novel two-stage Diffusion Probabilistic Model (Fig.~\ref{fig:method}), which synthesizes high-resolution MRIs conditionally-dependent on metadata (such as age). In the first stage, the MRI is turned into a quantized encoding derived from a `code book,' which is generated by Vector Quantization coupled with a Variational Autoencoder (VQ-VAE)~\cite{vqvae2017}. The second stage first uses a Generalized Linear Model~\cite{mcnamee2005regression} (GLM) to disentangle the quantized encoding into a metadata-specific encoding and a residual. After turning the residual into `R-Code' (via the code book of the first stage), we introduce a novel second model that learns the (categorical) distribution of R-Codes across real MRIs based on a discrete diffusion model similar to \cite{bond2022unleashing}. Unlike \cite{bond2022unleashing},  
the model consists of a conditional transformer, which learns long-range dependencies between the categorical elements of the R-Code using `masking' during training~\cite{he2022masked}. After completing training, \ourMethod synthesizes a new MRI by first generating (random) R-code, which is transformed into a residual. The residual is combined with the metadata-specific encoding, which is derived from random metadata values. The resulting quantized encoding is finally converted into an MRI using the decoder from the first stage. The individual components of \ourMethod are now reviewed in further detail.
\subsection{Stage I: Encoding MRIs via Fine-Grained Vector Quantization}\label{sec:fine-grained-VQ}
The VQ-VAE consists of an encoder \(\mathcal{E}\), a decoder \(\mathcal{D}\), and a codebook \(\mathcal{C}\). First, the encoder \(\mathcal{E}\) maps a 3D MRI \({x} \in \mathbb{R}^{D \times H\times W} \) to its lower-dimensional latent representation \mbox{\(\mathbf{z} := \mathcal{E}(x) \in \mathbb{R}^{d \times h\times w \times n_D}\)}. 
Next,  vector quantization discretizes the continuous latent space, such that an MRI can be represented by a set of indices (a.k.a., code) defined with respect to a codebook \\\mbox{\( \mathcal{C} := \{(k, e(k)): k = 1, 2, \ldots, N_{\mathcal{C}} \mbox{ and  } e(k) \in \mathbb{R}^{n_D}\}\)}. 
Each of the \(d \times h\times w \) vectors \(z \in \mathbb{R}^{n_D}\) from \(\mathbf{z}\) is paired with its closest representative \(e(k)\) from the codebook so that the corresponding code is 
 \( \mathcal{I}(z) := \text{arg min}_{k}(z-e(k)).\) Thus, \(e(\mathcal{I}(z))\) is the representative so that the quantization of the latent encoding may be formulated as \(Q(\mathbf{z}) = [e(\mathcal{I}(z))]_{z \dashv\,  \mathbf{z}}\), where $[ \cdot ]_{z \dashv\, \mathbf{z}}$ denotes the vectors $z$ defined according to $\mathbf{z}$. Note that the decoder network \(\mathcal{D}\) takes a quantized latent variable \(Q(\mathbf{z})\) and reconstructs the corresponding input image \(x\).


Applied to 3D (T1-weighted) MRIs, we discovered that the quantization based on the implementation by \cite{esser2021taming} leads to inaccurate approximations of \(\mathbf{z}\). We therefore propose a more fine-grained quantization, i.e., we first find its nearest representative \(e(\mathcal{I}(z))\) from the codebook for each vector z and then further search for an extra representative accounting for the difference \(z- e(\mathcal{I}(z))\). We do so by first reshaping the latent encoding $\mathbf{z}\in \mathbb{R}^{d \times h\times w \times n_D}$ so that $\mathbf{z}$ consists of twice as many vectors \(z \in \mathbb{R}^{\frac{n_D}{2}}\) with half the size, i.e., $\mathbf{z}\in \mathbb{R}^{2\times d \times h\times w \times \frac{n_D}{2}}$. For each of the vectors $z$, we then first determine the code \(\mathcal{I}(z)\) as done above (with \(e(k) \in \mathbb{R}^{\frac{n_D}{2}}\) also being half in size) and then compute the remainder \(\mathcal{I}(z- e(\mathcal{I}(z)))\)
so that the quantized encoding can be written as 
\begin{equation}\label{eq:qz}
    Q(z) := e(\mathcal{I}(z))+ e(\mathcal{I}(z- e(\mathcal{I}(z))).
\end{equation}
Our VQ-VAE is then completed by stacking two successive representations on top of each other (to create a vector of the original size) and feeding them into the decoder. 

\subsection{Stage II: Learning the distribution of R-Codes}
In the second stage, a sample is now encoded by the quantized representation \(Q(\mathbf{z})\) and metadata \(\mathbf{m}\). 
\subsubsection{Determining the R-Code}
Next, a GLM determines the metadata-specific matrix $\mathcal{P}$ so that each quantized encoding \(Q(\mathbf{z})\) is encoded by the metadata-specific component $\mathcal{P}\mathbf{m}$ and the residual component \(\mathbf{r}\), i.e., 
\[ Q(\mathbf{z}) = \mathcal{P}\mathbf{m} + \mathbf{r}.\] 
The matrix $\mathcal{P}$ is the least square solution across all training samples, which can be computed in closed form due to the relatively low dimension of the quantized encoding and assuming that the number of samples does not cause memory problems (which was not the case for our experiments). If one experiences memory issues, then one should use the  batch-wise solution as we proposed in ~\cite{lu2021metadata}. Once $\mathcal{P}$ is determined, \ourMethod computes (metadata-free) residual \(\mathbf{r}\) for each sample and applies the fine-grained VQ of Section~\ref{sec:fine-grained-VQ} to determine the corresponding R-Code $\mathcal{I}$.


\subsubsection{Mask Based Diffusion Model}
\label{MCDM}

\newcommand{\SET}{\widetilde{\mathcal{I}}}
\newcommand{\ID}{\mathbb{I}}

The distribution of R-Codes across real MRIs is now learned by a transformer architecture. While transformers can capture long-distance dependencies across the entries of an R-Code, their model complexity increases quadratically with the length of the R-Code. To make the training of the transformer efficient and scalable for our code (whose length in our experiment is over 4000 entries), we divide the R-Code $\mathcal{I}$ into \(N_{\mathcal{I}}\) subsets and train the transformer only on those sub codes. 

In particular, let `$q$' be the distribution of real R-Codes  and  $\mathcal{I} \sim q$ be a sample from that distribution with $N_{\mathcal{I}}$ partitions, i.e.,  $\left\{\mathcal{I}^{1}, \ldots, \mathcal{I}^{N_{\mathcal{I}}}\right\}$. For each \(n \in [1, N_{\mathcal{I}}]\), the transformer generates the target set $\SET := \mathcal{I}^n$ conditioned on the condition set $\SET'=:\mathcal{I}^{n-1}$. Note, for $n=1$ the condition set is empty. The transformer does so by training a Diffusion Probabilistic Model (DPM)~\cite{sohl2015deep,ho2020denoising}, which iterates between mapping 1) $\SET$ gradually to noise through a Forward Diffusion Process (FDP), and 2) noise back to $\SET$ through a Reverse Diffusion Process (RDP).

The FDP is formulated as a Markov chain with $\SET_t$ being a sample of the distribution conditioned on $\SET_0$. Unlike in the continuous case \cite{ho2020denoising}, one cannot add noise to $\SET_t$ by simply adding small random values to the entries of $\SET_t$ as each entry is a categorical variable is an index in the codebook. Therefore, minor changes in the index can have a large effect on the quantized encoding and thus the final synthesized MRI. Instead, we randomly mask out entries in $\SET_t$. Let  \(\mathbf{M}_t\) be the mask vector (of the same size as $\SET_t$) and each entry of \(\mathbf{M}_t\) be a binary variable with the probability \(\frac{1}{t}  \in [0, 1]\) being $1$ (and otherwise 0), then the noisy R-Code \(\SET_t\) at time step \(t\) is  
\begin{equation}
\label{eq:condition}
    \SET_t = \SET_{t-1} \odot \mathbf{M}_t.
\end{equation}
where `\(\odot\)' is the element-wise multiplication. 

Contrarily, the RDP goes through the entire Markov chain from time step $T$ to 0 in order to generate realistic $\SET_0$ from randomly masked out code  $\SET_T$ and conditioned on the already generated subset $\SET'$. It does so by simplifying the joint probability  $p(\SET_{0:T})=p(\SET_0, \ldots, \SET_T)$ to  
\begin{equation}
    p_\theta(\SET_{0:T}) :=  p(\SET_{T}) \prod_{t=1}^{T} p_{\theta}(\SET_{t-1}|\SET_t, \SET'), 
\label{eqn:conditionalP}
\end{equation}
with $p_{\theta}(\SET_{t-1}|\SET_t, \SET')$ being an approximation of the true posterior distribution. Now let $\mathcal{K}$ be a categorical distribution with \(N_{\mathcal{C}}\) categories, then 
\begin{equation}
    p_{\theta}(\SET_{t-1}|\SET_t, \SET') := \mathcal{K}(\SET_{t-1}; \mathbf{p} =logit(\epsilon_{\theta}(\SET_t, \SET', t))), 
\end{equation}
where the probability $\mathbf{p}$ is estimated by the logit of the output of a neural network $\epsilon_{\theta}(\cdot)$ with parameter $\theta$. RDP now learns to gradually unveils these masked values based on existing codes, i.e., those unmasked code from previous steps. Rather than directly approximating $\SET_{t-1}$, \ourMethod directly predicts the original code \(\SET_{0}\) thus reducing training stochasticity~\cite{ho2020denoising}. To do so, $\theta$ minimizes the cross-entropy between the predicted and the unmasked code entries, which is defined with respect to the expected value, i.e.,  
\begin{equation*}
- \mathbb{E}_{\SET_0 \sim q, t \in [0,\ldots, T] }\left[ \sum_{i \in [1, N_{\mathcal{I}}]} \log p_{\theta}(\SET_{0}[i]|\SET_t, \SET')\right].
\end{equation*}
Once minimized, the sampling of the code for the selected entries will be done according to Equation \ref{eqn:conditionalP}. Finally, in the last step (t=1) no entries will be masked out (i.e., no noise), which forces the model to create the entire subcode $\SET^n$.

\subsection{Synthesizing New MRI Conditioned on Metadata}
After training of the mask-based diffusion model has been completed, \ourMethod synthesizes MRI by first feeding empty subcodes $\left\{\mathcal{I}_T^{1}, \ldots, \mathcal{I}_T^{N_{\mathcal{I}}}\right\}$ in the reverse diffusion process. This process unmasks randomly selected entries in the subcode and fills them with `real values' by sampling from the learned distribution (of Stage II). Based on the partially filled subcode, more entries are unmasked and filled (see  Eq.~\eqref{eqn:conditionalP}). This process is repeated until all entries have values. Next, the generated subcodes $\left\{\mathcal{I}^{1}, \ldots, \mathcal{I}^{N_{\mathcal{I}}}\right\}$ are concatenated and transformed into a residual \(\mathbf{r}\). Next, random metadata values $\mathbf{m}$ are multiplied with the metadata-specific matrix $\mathcal{P}$ and the result is added to the residual to generate a quantized encoding $\mathbf{q}$. Finally, $\mathbf{q}$ is fed into the decoder $\mathcal{D}$ to generate a synthetic MRI $\mathcal{D}(\mathbf{q})$. 

Note, one could process each sub-code separately and at the end concatenate the corresponding synthetically generated subvolumes. However, this often leads to unsmooth boundaries between neighboring subvolumes, which is not an issue with our proposed strategy. Furthermore, using random metadata values enables our \ourMethod to create MRIs diversified with respect to the metadata. If one replaces the random values with specific ones (e.g. women of a certain age) then one can create counterfactual MRIs specific to that metadata (and not diverse).


%
%
\subsection{Implementation} 
We ran our implementation on an NVIDIA A100 GPU using the PyTorch 1.11.0 framework. With respect to Stage I, the encoder \(\mathcal{E}\) and decoder \(\mathcal{D}\) of the VQ-VAE are 3D CNN as in~\cite{esser2021taming}. The encoder and decoder each consist of four stages with the latent space being \(16=2^4\) times smaller than the original MRI. The codebook consists of  \(N_{\mathcal{C}}=1024\) representation units, each of which is a \(\frac{n_D}{2}=128\)D vector. This VQ-VAE model is trained using an Adam optimizer with 1000,000 iterations, a learning rate of $4.5e^{-6}$, and a batch size of 1.

With respect to Stage II, the R-code was divided into $N_{\mathcal{I}}=3$ sub-codes (48 slices each). Furthermore, the transformer model is constructed with a GPT architecture~\cite{radford2019gpt2} powered by blocks of multi-head self-attention mechanism, layer norm and
point-wise Multi-Layer Perceptrons. The transformer model is trained for 500,000 iterations with a batch size 24 and using the Adam optimizer with a learning rate of $10^{-4}$. 

\subsection{Materials} \label{sec:materials}
Our dataset consisted of 4296 MRIs (from 1,236 normal controls) pooled from the ADNI, NCANDA, and SRI data sets (Table \ref{tab:demographics}). The ADNI set~\cite{petersen2010alzheimer} included 1215 MRIs (of 342 controls from ADNI 1, 2, 3 and GO) that successfully passed through our preprocessing pipeline. 
The NCANDA set consisted of 2285 MRIs from the 621 participants of NCANDA$\_$PUBLIC$\_$6Y$\_$STRUCTURAL$\_$V01  ~\cite{ouyang2022self} that reported no-to-low alcohol drinking in the past year (prior to MRI acquisition) according to the adjusted Cahalan score. The SRI data included all 796 MRIs (from 273 normal controls) acquired by the study. 

As in ~\cite{zhao2021longitudinal,wei2023cDPM}, each scan was denoised, homogeneity-corrected, skull-stripped, and affinely aligned to a template resulting in an MRI of 1mm$^3$ voxel resolution and a spatial resolution of \(144\times 176\times 144\). The intensity values of the MRI were normalized between 0 and 1. 

We split the dataset into a training set and a replication set. The training set consisted of 3996 MRIs from 1,236 normal controls. The replication set consisted of 150 males and 150 females from those 1,236 subjects. The 300 MRIs in the replication set were from a different visit than the MRIs in the training set. Specifically, we separated the age range (13 to 90.9 years) into 10 groups. For each age group, we randomly selected 15 males and 15 females, who had multiple longitudinal visits. For each subject, the MRI corresponding to the age was selected and any other MRIs of that subject were discarded to avoid subjects being part of multiple age groups. The remaining 3996 MRIs were used for training \ourMethod.


For predicting brain age, we chose an independent dataset collected at UCSF consisting of the cross-sectional MRIs of all 486 subjects~\cite{zhang2022multi}. The subjects were divided into 4 cohorts:  156 Controls, 37 PLWH, 148 MCI, and 145 HAND. 


\section*{Data Availability}
This study involved two public datasets (ADNI and NCANDA). ADNI data was obtained from the public data releases ADNI 1, 2, 3 and GO accessible via \url{https://adni.loni.usc.edu/}. The NCANDA data is part of the public data release NCANDA\_PUBLIC\_6Y\_STRUCTURAL\_V01 \cite{ouyang2022self}, whose collection and distribution were supported by NIH funding AA021681, AA021690, AA021691, AA021692, AA021695, AA021696, AA021697 and is distributed according to the NCANDA Data Distribution agreement https://www.niaaa.nih.gov/ncanda-data-distribution-agreement.


\section*{Code Availability}
The deep learning model \ourMethod was written in Python 3.7 and PyTorch 1.11.0, and is available to reviewers via the service provided by Nature on Code Ocean. Once the article is accepted for publication, the code will also be publicly available at  \url{https://github.com/xiaoiker/Meta_DPM}. The code for analyzing the results was written in Python using standard, open-source Python libraries.

\section*{Acknowledgement}
This work was partly supported by funding from the National Institute of Health (MH113406, DA057567, AA021697, AA017347, AA010723, AA005965, and AA028840), the DGIST R\&D program of the Ministry of Science and ICT of KOREA (22-KUJoint-02), Stanford School of Medicine Department of Psychiatry and Behavioral Sciences Faculty Development and Leadership Award, and by the Stanford HAI Google Cloud Credit.

\section*{Author Contribution}
WP, QZ, EA, and KMP shaped the core ideas and designed the experiments. QZ and KMP prepared the data for analysis. WP developed the methodology, worked out the technical details, and analyzed the data. TB contributed to data preparation and analysis. WP and QZ drafted the initial manuscript. All authors contributed to editing. KMP and QZ directed and supervised the project.
\bibliographystyle{unsrt}
\bibliography{refs}

\end{document}